\documentclass[lettersize,journal]{IEEEtran}
\usepackage{amsmath,amsfonts}
\usepackage{array}
\usepackage[caption=false,font=normalsize,labelfont=sf,textfont=sf]{subfig}
\usepackage{textcomp}
\usepackage{stfloats}
\usepackage{url}
\usepackage{verbatim}
\usepackage{graphicx}
\usepackage{cite}
\usepackage[colorlinks,linkcolor=black,anchorcolor=blue,citecolor=blue]{hyperref}
\usepackage[ruled, vlined, linesnumbered]{algorithm2e}

\usepackage{listings}
\usepackage{xcolor}
\usepackage{rotating}
\usepackage{multirow}
\definecolor{codegreen}{rgb}{0,0.6,0}
\definecolor{codegray}{rgb}{0.5,0.5,0.5}
\definecolor{codepurple}{rgb}{0.58,0,0.82}
\definecolor{backcolour}{rgb}{0.95,0.95,0.92}

\usepackage{multirow}
\usepackage{diagbox}
\usepackage{makecell}
\usepackage{tabularx}
\usepackage{adjustbox}
\usepackage{utfsym}
\usepackage{booktabs}

\newcommand{\mypara}[1]{\smallskip\noindent\textbf{#1.} \xspace}
\usepackage{cite}
\usepackage{amsmath,amssymb,amsfonts}
\usepackage{algorithmic}
\usepackage{graphicx}
\usepackage{textcomp}
\usepackage{xcolor}
\usepackage{multirow}
\usepackage{caption}
\usepackage{float} 
\usepackage{amssymb}
\usepackage{diagbox}
\usepackage{makecell}
\usepackage{threeparttable}
\usepackage{fontawesome}
\usepackage{tabularx}
\usepackage{booktabs}
\usepackage{bbding}
\usepackage{pifont}

\lstdefinestyle{mystyle}{
    backgroundcolor=\color{backcolour},   
    commentstyle=\color{codegreen},
    keywordstyle=\color{magenta},
    numberstyle=\tiny\color{codegray},
    stringstyle=\color{codepurple},
    basicstyle=\ttfamily\footnotesize,
    breakatwhitespace=false,         
    breaklines=true,                 
    captionpos=b,                    
    keepspaces=true,                 
    numbers=left,                    
    numbersep=5pt,                  
    showspaces=false,                
    showstringspaces=false,
    showtabs=false,                  
    tabsize=2
}

\begin{document}

\title{CIBPU: A Conflict-Invisible Secure Branch Prediction Unit}

\author{Zhe Zhou, Fei Tong, Hongyu Wang, Xiaoyu Cheng, Fang Jiang, Zhikun Zhang, Yuxing Mao
        
        \IEEEcompsocitemizethanks{\IEEEcompsocthanksitem This work is supported in part by the National Natural Science Foundation of China (No. 61971131), and in part by ``Zhishan'' Scholars Programs of Southeast University. (Corresponding Author: Fei Tong)
			\IEEEcompsocthanksitem Z. Zhou, F. Tong, X. Cheng and F. Jiang are with the School of Cyber Science and Engineering, Southeast University, Nanjing, Jiangsu, China (\{220215213, ftong, xiaoyu\_cheng, 230239271\}@seu.edu.cn).
			\IEEEcompsocthanksitem F. Tong is also with the Purple Mountain Laboratories, Nanjing, Jiangsu, China, and the Jiangsu Province Engineering Research Center of Security for Ubiquitous Network, China.
			\IEEEcompsocthanksitem H. Wang is with the State Key Laboratory of Power Transmission Equipment \& System Security and New Technology, Chongqing University, and Wiscom System Co., LTD (awang@wiscom.com.cn).	
			\IEEEcompsocthanksitem Z. Zhang is with Stanford University (zhikun@stanford.edu).
			\IEEEcompsocthanksitem Y. Mao is with the State Key Laboratory of Power Transmission Equipment \& System Security and New Technology, Chongqing University (myx@cqu.edu.cn).
		}
}

\maketitle
\begin{abstract}
Previous schemes for designing secure branch prediction unit (SBPU) based on physical isolation can only offer limited security and significantly affect BPU's prediction capability, leading to prominent performance degradation. Moreover, encryption-based SBPU schemes based on periodic key re-randomization have the risk of being compromised by advanced attack algorithms, and the performance overhead is also considerable. To this end, this paper proposes a conflict-invisible SBPU (CIBPU). CIBPU employs redundant storage design, load-aware indexing, and replacement design, as well as an encryption mechanism without requiring periodic key updates, to prevent attackers' perception of branch conflicts. We provide a thorough security analysis, which shows that CIBPU achieves strong security throughout the BPU's lifecycle. We implement CIBPU in a RISC-V core model in gem5. The experimental results show that CIBPU causes an average performance overhead of only 1.12\%--2.20\% with acceptable hardware storage overhead, which is the lowest among the state-of-the-art SBPU schemes. CIBPU has also been implemented in the open-source RISC-V core, SonicBOOM, which is then burned onto an FPGA board. The evaluation based on the board shows an average performance degradation of 2.01\%, which is approximately consistent with the result obtained in gem5.



\end{abstract}

\section{Introduction}\label{sec:Introduction}

\IEEEPARstart{B}{ranch} prediction is an advanced method to improve the pipeline stall problem caused by branch instructions~\cite{mcfarling1986reducing}, with which, the CPU can judge the direction and jump to the address of the program branch, accelerating the running speed.
\textbf{\textit{Branch prediction unit} (BPU)} is a component of CPU designed to support branch prediction.
While BPU has brought huge performance gains to modern CPUs, it has also introduced security problems. 
Early studies have revealed the BPU side-channel vulnerabilities, leading to information leakage. (e.g., SBPA~\cite{aciiccmez2007power}). 
Recently, researchers have unveiled the significant threat posed by transient execution attacks that leverage branch predictors to CPU security. These attacks utilize BPU as covert channels~\cite{evtyushkin2018branchscope,huo2020bluethunder,lee2017inferring}, or maliciously train BPU to trigger transient executions that leak sensitive information from arbitrary memory locations through other covert channels~\cite{kocher2019spectre,koruyeh2018spectre,canella2019systematic}.


The fundamental reasons for the existence of these attacks can be attributed to two factors. First, the individual threads of the processor typically share all the resources of BPU, which will cause branch conflicts between different threads storing information in BPU. 
As such, the adversaries can exploit the \textbf{\textit{simultaneous multi-threading} (SMT)}~\cite{nemirovsky2013simultaneous} techniques to actively create branch conflicts.
Second, BPU entries are usually indexed by the values obtained from compressing the PC and branch historic information. 
These two design features can result in branch conflicts. 
Concretely, the method of how branch prediction information is stored in BPU is similar to cache. 
An attacker can use a similar attack method as in cache (e.g., Prime+Probe~\cite{osvik2006cache}) to evict the victim entry from BPU if the attacker's virtual addresses are mapped to the same set as the victim (such an event is called a set-conflict event). 
The attacker uses this eviction to monitor the victim's access patterns and thus infer critical information. 
On the other hand, in a Spectre attack, the attacker can inject malicious branch targets at specific locations of BPU and wait for the victim's access to the same locations, resulting in the transient execution of the malicious branches. 
The transient information generated by these malicious branches can be leaked through other micro-architectural covert channels.


\begin{scriptsize}
\begin{table*}[!t]
\caption{Comparison of CIBPU with the state-of-the-art SBPU schemes.}
\centering
\begin{threeparttable}
\begin{tabular}{ccccccc}
\midrule[0.8pt]
\textbf{SBPU Schemes}
& \makecell[c]{\textbf{Protection} \\ \textbf{Principle}}
& \makecell[c]{\textbf{Performance Overhead}\tnote{1}}
& \textbf{\makecell[c]{Hardware \\ Storage Cost}}
& \textbf{\makecell[c]{Without \\ Re-randomization?}}
& \textbf{\makecell[c]{Single-Thread  \\ Security}}
& \textbf{\makecell[c]{SMT \\  Security}}
\\ 
\midrule[0.8pt]

Flush~\cite{evtyushkin2016understanding}
& Isolation
& \makecell[l]{Single-threaded: 6.63\%--10.2\% \\ SMT-2: /}
& /
& {\ding{51}}
& \textcolor{teal}{\ding{51}} 
& \textcolor{red}{\ding{55}} 
\\ 
\midrule

Partition~\cite{evtyushkin2016understanding}
& Isolation
& \makecell[l]{Single-threaded: 5.50\%--6.72\% \\ SMT-2: 6.70\%}
& /
& {\ding{51}}
& \textcolor{red}{\ding{55}} 
& \textcolor{teal}{\ding{51}} 
\\ 
\midrule

Replication~\cite{vougioukas2019brb} 
& Isolation
& \makecell[l]{Single-threaded: / \\ SMT-2: -0.70\%}
& \makecell[l]{SMT-2: 212.3KB\\ SMT-8: 1.04MB}
& {\ding{51}}
& \textcolor{red}{\ding{55}} 
& \textcolor{teal}{\ding{51}} 
\\
\midrule

HyBP~\cite{zhao2022hybp}
& Encryption
& \makecell[l]{Single-threaded: 1.73\%--5.69\% \\ SMT-2: 3.52\%}
& \makecell[l]{SMT-2: 39.6KB\\ SMT-8: 197.5KB}
& {\ding{55}}\tnote{2}  
& \textcolor{teal}{\ding{51}} 
& \textcolor{teal}{\ding{51}} 
\\ 
\midrule

STBPU~\cite{zhang2022stbpu} 
& Encryption
& \makecell[l]{Single-threaded: 2.82\%--3.94\% \\ SMT-2: 3.33\%}
& \makecell[c]{/}
& \ding{55}\tnote{3}
& \textcolor{teal}{\ding{51}} 
& \textcolor{teal}{\ding{51}} 
\\ 
\midrule

\textbf{CIBPU (ours)} 
& \textbf{Encryption}
& \makecell[l]{\textbf{Single-threaded: 1.12\%--2.20\%} \\ \textbf{SMT-2: 2.12\%}}
& \makecell[l]{\textbf{SMT-2: 59.3KB}\\ \textbf{SMT-8: 59.3KB}}
& \ding{51}
& \textcolor{teal}{\ding{51}} 
& \textcolor{teal}{\ding{51}}
\\  
\midrule[0.8pt]
\end{tabular}

\vspace{0.3em}
 \begin{tablenotes}
        \item[1] The testing was conducted in gem5. In the default configuration, the context switch frequency is set to 1K--10K switches per second for single-threaded mode, and 3000 switches per second for SMT-2 (The number of threads running at the same time is 2) mode.
        \item[2] HyBP re-randomizes the keys during context switches.
        \item[3] STBPU re-randomizes the keys when the number of detected malicious BPU events reaches a threshold.
        \item[]\ding{51}: Yes. \ding{55}: No. \textcolor{teal}{\ding{51}}: The security is supported. 
        \textcolor{red}{\ding{55}}: The security is not supported.
\end{tablenotes}
\end{threeparttable}
\label{tab:1}

\end{table*}
\end{scriptsize}

\mypara{Existing solutions}
In recent years, only a few studies focus on the design of \textbf{\textit{secure BPU} (SBPU)}, which can be classified into two categories: Providing physical isolation for BPU resources~\cite{evtyushkin2016understanding,vougioukas2019brb}, and deploying encryption mechanism to BPU~\cite{lee2020securing,zhao2021lightweight,zhao2022hybp,grayson2020evolution,zhang2022stbpu}. 
Concretely, isolation-based SBPUs implement thread- or privilege-level isolation for BPU, but they will cause large performance degradation since the BPU resources shared among threads for reuse are reduced. 
Furthermore, BPU under this type of protection scheme still has security vulnerabilities~\cite{zhao2022hybp}. 
On the other hand, encryption-based SBPUs encrypt the index and content of BPU with a randomly generated key for each thread, which needs to be updated periodically (thousands of times per second) through re-randomization to guarantee BPU security. 
Therefore, they require additional hardware storage to store thread-specific keys. 
In addition, high-frequency key re-randomization may greatly degrade processor performance. 
Recently, Bourgeat et al. ~\cite{bourgeat2020casa} showed that the periodic key re-randomization itself may gradually accumulate leaks over multiple periods of time, which could be utilized by a more advanced attacks.




\mypara{Our proposal}
An efficient SBPU should solve the root cause of attacks, i.e., branch conflicts. However, completely eliminating branch conflicts is challenging. 
In this paper, we propose \textbf{\textit{conflict-invisible secure BPU} (CIBPU)}, a practical SBPU mechanism aiming to protect against conflict-based attacks by concealing branch conflicts from attackers. 
CIBPU also employs encryption techniques without requiring periodic key re-randomization to protect the data in \textbf{\textit{pattern history table} (PHT)} and \textbf{\textit{branch target buffer} (BTB)}, which are the two of the most important components in BPU. 
To prevent attackers from perceiving the branch conflicts within BPU throughout its lifecycle, two indexing and replacement strategies for PHT and BTB are designed, namely CIPHT and CIBTB.

\mypara{Contributions} In summary, our contributions are:

\begin{itemize}
    \item We propose CIBPU, an SBPU scheme that effectively prevents conflict-based BPU attacks. CIBPU includes CIPHT and CIBTB, which are designed to provide protection measures for two critical components of BPU, namely PHT and BTB, respectively.
    \item We provide a thorough analysis on the security of CIBPU, which shows that CIBPU ensures a strong security throughout its lifecycle without requiring key re-randomization.
    \item We implement CIBPU in a RISC-V core model in gem5, and evaluate it in comparison with the state-of-the-art SBPU schemes, as shown in \autoref{tab:1}. By executing the SPEC CPU 2017 benchmarks (SPEC2017~\cite{SPECCPU2017}), CIBPU shows the lowest performance degradation. 
    \item We also implement CIBPU in the open-source core, SonicBOOM~\cite{SonicBOOM3rdGeneration2020}. The core is then burned onto an FPGA evaluation board, based on which a Linux operating system is run to execute SPEC2017 for performance overhead evaluation. The obtained results are approximately consistent with those obtained through gem5~\cite{gem5simulatorsystem}.

\end{itemize}



\section{Background}\label{sec:back}

\begin{figure*}[t]
    \centering
    \includegraphics[width=.9\linewidth]{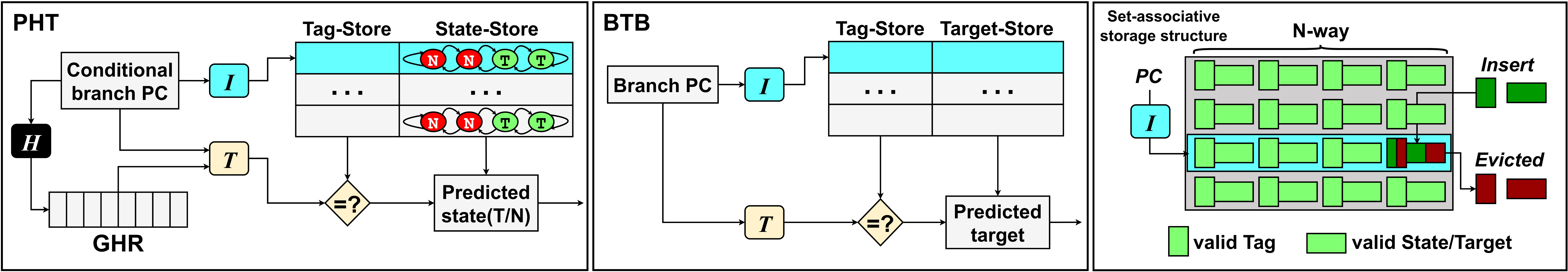}
    \vspace{-0em}
    \caption{The internal structure of the conventional PHT and BTB.} 
    \label{figbackground}
    \vspace{-1em}
\end{figure*}

\subsection{Branch Prediction Unit}
\label{subsec:bpu}

In modern CPUs, BPU typically consists of PHT and BTB. 
The internal structure of the conventional PHT and BTB is illustrated in \autoref{figbackground}. PHT and BTB have a structure similar to cache, where they utilize a designed hash function on the \textbf{\textit{program counter} (PC)} of branch-type instructions to construct index values (denoted as \textbf{\textit{I}}) and tag values (denoted as \textbf{\textit{T}}). 
Concretely, to reduce branch conflicts, PHT introduces an additional shift register known as the global \textbf{\textit{history register} (GHR)} to store the history of branch jumps. 
PC is transformed by a specified mapping function to construct the index value  \textbf{\textit{H}} of GHR. 
The stored information in GHR also contributes to the construction of the tag value. 
Both PHT and BTB use the index value \textbf{\textit{I}} and tag value \textbf{\textit{T}} to access entries and determine if there is a hit.

\mypara{PHT entries} 
The PHT entries consist of two components, namely \textit{Tag-Store} and \textit{State-Store}. 
The Tag-Store stores the tag values used for hit detection, while the State-Store contains a 2-bit counter that represents the current state of the branch prediction. 
A count value of 2 or 3 indicates a predicted branch taken, while a count value of 0 or 1 indicates a predicted branch not taken. The 2-bit counter is updated as follows: It increments by 1 when the predicted branch is taken and decrements by 1 when the branch is not taken. 
The counter remains unchanged when it reaches the upper or lower limit.  

\mypara{BTB entries} 
The BTB entries have a structure more similar to a cache. 
They consist of two components, namely \textbf{\textit{Tag-Store}} and \textbf{\textit{Target-Store}}. 
The Tag-Store is used to store the tag values for hit detection, while the Target-Store stores the addresses of the branches to be taken.

\mypara{Set-associative storage structure} 
As shown in \autoref{figbackground}, to improve the hit rate of BPU, the storage structure of PHT and BTB entries adopts the commonly used set-associative technique as in caches. 
PC indexes a set of entries, and then the tag values of all the entries in this set are compared with \textbf{\textit{T}}. If no match is found, a miss occurs, and one of the entries in the set needs to be selected and updated with a new pair of tags and Target/State (the original pair is evicted). 
The update strategy is typically based on the least-recently-used replacement algorithm or random replacement algorithm. 
Both algorithms operate within the set scope, meaning they select one entry within the set to evict for the newly inserted entry. 
This characteristic leads to the second type of BPU conflict-based attacks to be introduced below. 

\subsection{BPU Attacks} 

In \autoref{sec:Introduction}, we classify attacks against BPU into two categories based on the locations where information leakage occurs. 
In this section, we reclassify attacks on BPU based on the behavior of attackers.


There are two BPU features that are present in almost all CPUs, and they make branch conflict-based attacks possible. 
Firstly, the BPU data structures are shared among all software running on the CPU core, enabling branch conflicts between different processors. 
Secondly, the BPU operates on compressed virtual addresses. For example, out of the 48 bits in a branch virtual address, only a part of the bits are used, and these bits are further compressed. This allows conflicts to occur within the same virtual address space, such as conflicts between branches in the kernel and user processes. 
The determinism of the BPU makes it possible for attackers to trigger branch conflict in a controlled manner. 
Current research has developed two types of branch conflict-based attacks. In \autoref{sec:Security}, we describe the security analysis for these attacks.

\mypara{Reuse-based attacks} 
In the PHT and BTB structure, the attacking process can employ malicious training to influence the (speculative) execution of the victim process, which in turn can lead to or enhance information leakage. 
Different programs can directly access shared resources. When one process sets the branch direction or target information, it may impact another process, resulting in a branch conflict. 
We refer to attacks that exploit this characteristic as \textbf{\textit{reuse-based attacks}}. 

\mypara{Eviction-based attacks} 
In the BTB structure, a cache-like storage structure, attacks similar to cache eviction attacks can be launched. 
The attacker detects whether there is a contention in the corresponding entry of the branch predictor table to gain insights into the execution behavior of the victim's target branch. 
This represents another form of branch conflict. 
We refer to the attacks that leverage this characteristic as \textbf{\textit{eviction-based attacks}}. 

Thzze goal of CIBPU is to provide a secure and reliable BPU design that effectively defends against branch conflict-based attacks, thus safeguarding the security of systems and user data.

\section{Threat Model}

\mypara{Attacker's capabilities}
This paper considers the scenario where the attacker thread and the victim thread run on the same processor core, and the attacker has knowledge of the victim's source code and address layout. In order to exercise fine-grained control over the execution of victim processes, attackers have the capability to slow down the running speed of these processes, which is also a prerequisite for certain attacks~\cite{huo2020bluethunder,evtyushkin2018branchscope,lee2017inferring}.

\begin{table}[!t]
\caption{The conflict-based BPU attacks and their categorization that CIBPU can defend against.\label{tab:attacks}}
\centering
\begin{tabular}{ccc}
\midrule[0.8pt]
        \makecell[c]{\textbf{Conflict-based} \\ \textbf{attacks}} & \makecell[c]{\textbf{BPU side/covert-} \\ \textbf{channel attacks}} & \makecell[c]{\textbf{Spectre attacks} \\ \textbf{using BPU}} \\
\midrule[0.8pt]
        \makecell[c]{\textbf{Reuse-based} \\  \textbf{attacks}} & \makecell[c]{BranchScope~\cite{evtyushkin2018branchscope} \\ Bluethunder~\cite{huo2020bluethunder} \\ Branch Shadowing~\cite{lee2017inferring} \\  Spectre-NDA-BTB~\cite{weisse2019nda}}   & \makecell[c]{Spectre-BHB~\cite{barberis2022branch} \\ Spectre-BTB~\cite{kocher2019spectre}}\\
\midrule
        \makecell[c]{\textbf{Eviction-based} \\ \textbf{attacks}} & \makecell[c]{SBPA~\cite{aciiccmez2007power} \\ JUMP~\cite{evtyushkin2016jump} \\ Spectre-NDA-BTB~\cite{weisse2019nda}}  & \makecell[c]{Spectre-BHB~\cite{barberis2022branch} \\ Spectre-BTB~\cite{kocher2019spectre}}\\
\midrule[0.8pt]
\end{tabular}
\vspace{-20pt}
\end{table}

\mypara{Defense goal}
This paper aims to defend against conflict-based attacks related to BPU, including reuse-based attacks (e.g., Spectre-BTB~\cite{kocher2019spectre} and BranchScope~\cite{evtyushkin2018branchscope}) and eviction-based attacks based on BTB eviction (e.g., Spectre-BTB~\cite{kocher2019spectre} and JUMP~\cite{evtyushkin2016jump}). The complete list of these attacks and further categorization is shown in \autoref{tab:attacks}. CIBPU focuses on attacks where the attacker and victim are not on the same thread, which is the most common potential attacker scenario. The attacker thread and the victim thread can run in SMT mode or in a round-robin fashion. CIBPU does not consider attacks in which the attacker and victim are on the same thread, such as Specter-PHT~\cite{kocher2019spectre} and Malicious Trojan~\cite{zhang2020exploring}. Such attacks usually require careful verification of application code or rely only on link libraries provided by trusted manufacturers. Otherwise, A defense method based on selective speculative execution~\cite{yu2019speculative,weisse2019nda} needs to be implemented in the CPU to completely prevent this type of attack. In addition, CIBPU focuses on mitigating side/covert-channel attacks of the branch predictor itself and speculative covert-channel attacks that exploit the branch predictor. Traditional side-channel attacks on other microarchitectural components (such as cache~\cite{bernstein2005cache}, TLB~\cite{gras2018translation}) are beyond the scope.



\section{Our Proposal: CIBPU}\label{sec:Design}

\subsection{Overview}

Based on the analysis of existing SBPU mechanisms in \autoref{sec:Introduction}, an efficient and practical SBPU is expected to have the following features:
\begin{itemize}
    \item Versatile and can be easily applied to existing BPUs.
    \item Limited performance and hardware storage overhead.
    \item Provable and sufficient security.
\end{itemize}

To achieve the aforementioned features, we propose CIBPU. As shown in \autoref{fig:enc}, \autoref{fig:PHT}, and \autoref{fig:BTB}, the design details of CIBPU can be divided into three parts: Two-level encryption, CIPHT, and CIBTB.


\mypara{Part 1: Two-level Encryption} As shown in the \autoref{fig:enc}, the value of index and stored contents of PHT and BTB in CIBPU are encrypted, which we refer to as two-level encryption. The details of the encryption method are presented in \autoref{Enc}. Moreover, the encryption method have been modified respectively due to the unique designs of CIPHT and CIBTB, which are demonstrated respectively in \autoref{PHT} and \autoref{BTB}.

\mypara{Part 2: CIPHT} As shown in the \autoref{fig:PHT}, CIPHT replicates PHT entries three times and divides them into three \textit{skews}\footnote{\textit{Skew}: In storage structures like cache and branch predictors, \textit{skew} is a grouping method used to allocate data to different regions (i.e., \textit{skews}) based on various rules~\cite{michaud1996skewed}.}, which are encrypted with different keys for protection. The specific design details are presented in \autoref{PHT}.

\mypara{Part 3: CIBTB} As shown in the \autoref{fig:BTB}, CIBTB decouples the tag storage and address storage in the BTB entries, dividing the tag storage part into two skews and encrypting the index values and tag values of each skew with different keys. Additionally, we have designed a load-aware indexing and replacement algorithm tailored to this unique hardware structure, which is detailed in \autoref{BTB}.


\subsection{Two-level Encryption}
\label{Enc}

\begin{figure}[t]
    \centering
    \includegraphics[width=0.7\linewidth]{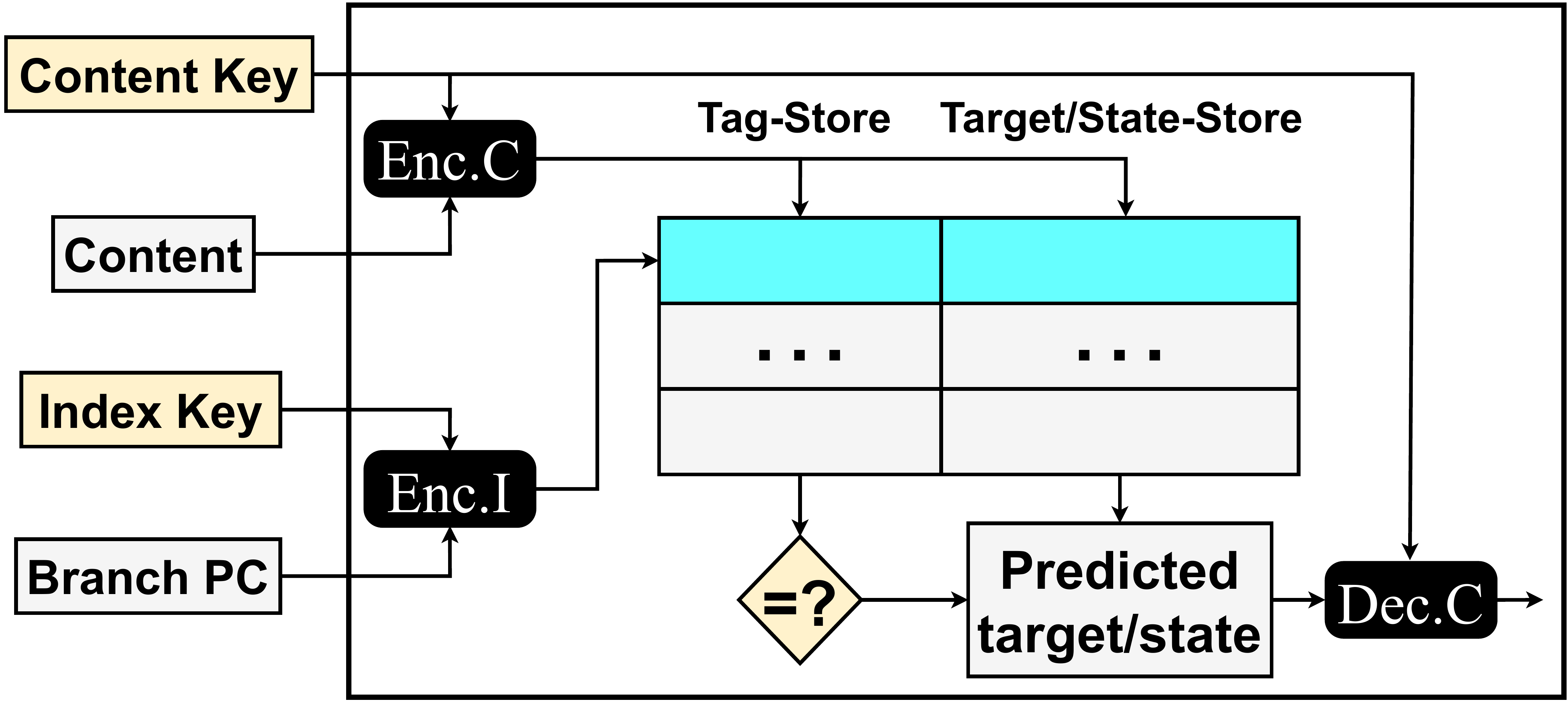}
    \vspace{-5pt}
    \caption{Overview of index encryption and content encryption mechanisms.} 
    \label{fig:enc}
    \vspace{-15pt}
\end{figure}

The purpose of encryption is to shield the mutual perception of BPU resources among different threads, greatly impeding the possibility of reuse-based attacks. This ensures that even in the presence of shared storage, when one process executes a branch, it will not obtain the storage location and content of another process due to the differences in their encryption keys. \autoref{fig:enc} illustrates the fundamental encryption mechanism of CIPHT or CIBTB. CIPHT or CIBTB encrypts the index and content (tag or stored branch information) of PHT or BTB separately (corresponding to \textbf{Enc.I} and \textbf{Enc.C} in \autoref{fig:enc}, respectively). When a BPU hit happens, the same key is used to decrypt the content and restore the information (\textbf{Dec.C} in \autoref{fig:enc}). The specific \textbf{Enc.I} and \textbf{Enc.C} designs for CIPHT and specific \textbf{Enc.I} design for CIBTB will be described in detail in the following subsections.


\subsection{CIPHT}
\label{PHT}

\begin{figure}[t]
    \centering
    \includegraphics[width=0.7\linewidth]{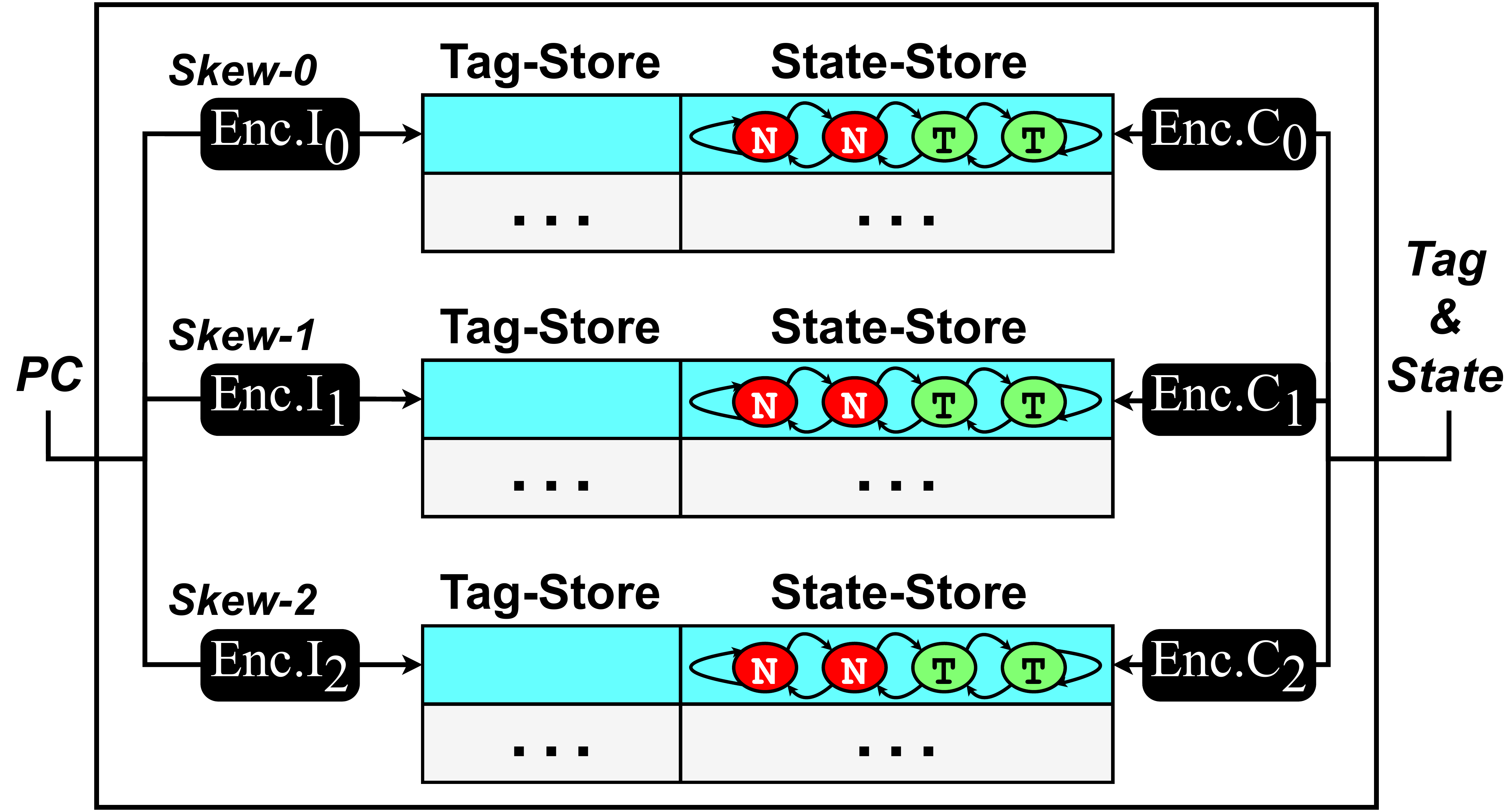}
    \vspace{-5pt}
    \caption{Overview of design of CIPHT.} 
    \label{fig:PHT}
    \vspace{-10pt}
\end{figure}

As shown in \autoref{fig:PHT}, we replicate the original entries of PHT into three distinct skews. As introduced in \autoref{subsec:bpu}, the \textit{State-Store} portion of every PHT entry only stores 2 bits of state information. Therefore, replicating PHT entries does not cause excessive space overhead. When querying PHT, each entry in each skew is indexed by encrypted PC: Enc.I$_0$/Enc.I$_1$/Enc.I$_2$. Also, the stored content (tags and states) in each skew is encrypted by Enc.C$_0$/Enc.C$_1$/Enc.C$_2$.
In this way, we achieve the same effect as increasing the encrypted index and tag bit count, making the probability of an attacker reusing the PHT state be almost zero, which will be analyzed in detail in \autoref{sec:Security}.

Under this strategy, there are slight changes to the querying and replacement mechanisms of PHT. Specifically, when an instruction's PC hits all three tables, it is considered as a PHT hit, and the subsequent prediction operations are executed. For convenience, we select the state information stored in any one of the tables as the chosen state. Conversely, when a PHT miss occurs, we perform replacement operations on all three tables. The behavior of the three tables is completely unified under our mechanism.

\subsection{CIBTB}\label{BTB}

CIBTB consists of three key components. Firstly, similar to the storage structure in V-way cache proposed in ~\cite{qureshi2005v}, we decouple Tag-Store and Target-Store to reduce branch conflicts by providing additional invalid tag options . Secondly, we divide all Tag-Stores into two skews and propose a load-balancing index algorithm to index the entries of each skew using two different keys. Lastly, we propose a load-balancing replacement algorithm, which selects the replacement data from both skews to achieve global eviction. The proposed two algorithms ensure that any intra-group eviction does not occur throughout the system's lifecycle. The three components are detailed below.

\begin{figure}[t]
    \centering
    \includegraphics[width=0.9\linewidth]{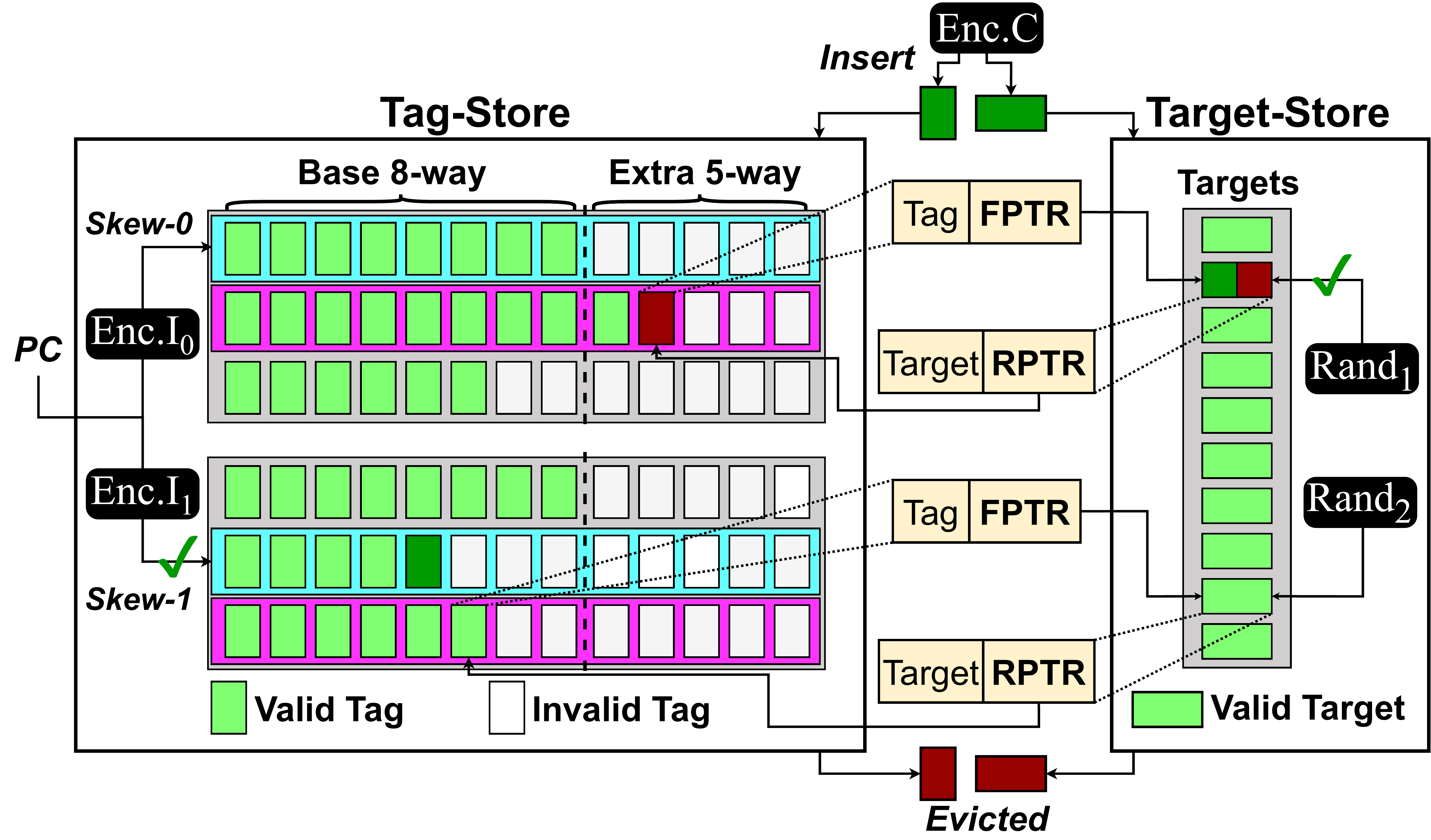}
    \vspace{-.5em}
    \caption{Overview of design of CIBTB.} 
    \label{fig:BTB}
    \vspace{-1em}
\end{figure}

\mypara{Decoupled Tag-Store and Target-Store structure} 
\autoref{fig:BTB} illustrates the organization of Tag-Store and Target-Store in CIBTB. The number of Tag-Stores exceeds the number of Target-Stores. Each Tag-Store has a \textbf{\textit{forward pointer} (FPTR)} that can point to any Target-Store, while each Target-Store has a \textbf{\textit{backward pointer} (RPTR)} that can point to any Tag-Store. This design follows the inherent structure of V-way cache. 

Within this structure, two types of evictions can be observed. The first type, named \textbf{\textit{secure eviction} (SE)}, occurs when a new branch type instruction evicts a branch from outside the set it was originally indexed to. The second type, named \textbf{\textit{dangerous eviction} (DE)}, happens when a new branch information evicts a branch from within the set it was indexed to. This situation arises when the valid tag count in the set indexed by the branch type instruction reaches its capacity, potentially resulting in information leakage. Inspired by MIRAGE~\cite{saileshwar2021mirage}, we introduce additional invalid tags for each set. This effectively reduces the occurrence rate of DE.

\mypara{Load-balancing index strategy for two skews} 
To reduce intra-set evictions during the process of BPU indexing, where sets always contain invalid tags, we propose a load-balancing index algorithm and divide all sets into two skews. This approach aims to increase the probability of avoiding intra-set evictions. The complete indexing process is illustrated in \autoref{fig:BTB} and described in ~\autoref{a1}. Specifically, we use two different keys to map the PC to a set within each skew. The encryption mechanism ensures that attackers cannot arbitrarily consume invalid tags within a set to cause DE to occur.

\begin{algorithm}[!t]
\footnotesize
\caption{Load-balancing index algorithm}\label{a1}
\LinesNumbered
\KwIn{{\tt Key$_0$}, {\tt Key$_1$}, {\tt Key$_\text{c}$}, {\tt PC}, \textit{Sets}}
\KwOut{{\tt BTB\_State}, {\tt Final\_Set}}

{\tt Enc.I$_0$} $\gets$ {\tt PC} $\oplus$ {\tt Key$_0$}; {\tt Enc.I$_1$} $\gets$ {\tt PC} $\oplus$ {\tt Key$_1$};

{\tt Set$_0$} $\gets$ \textit{Sets}$\text{[}${\tt Enc.I$_0$}$\text{]}$; {\tt Set$_1$} $\gets$ \textit{Sets}$\text{[}${\tt Enc.I$_1$}$\text{]}$;

{\tt tag $\gets$ {\tt PC} $\oplus$ {\tt Key$_\text{c}$}};


{\tt N$_0$} $\gets$ Number of valid tags in {\tt Set$_0$};

{\tt N$_1$} $\gets$ Number of valid tags in {\tt Set$_1$};

\uIf{{\tt tag} {\rm hits} {\tt Set$_0$} {\rm or} {\tt Set$_1$}}{
    {\tt BTB\_State} $\gets$ {\textit{Hit}};\\
    return;
}\uElseIf{ {\tt N$_0$} $<=$ {\tt N$_1$} }{
    {\tt BTB\_State} $\gets$ {\textit{Miss}}; {\tt Final\_Set} $\gets$ {\tt Set$_0$};
}\Else{
    {\tt {BTB\_State}} $\gets$ {\textit{Miss}}; {\tt Final\_Set} $\gets$ {\tt Set$_1$};
}

\end{algorithm}

\mypara{Load-balancing replacement strategy} 
When a BTB miss occurs, two targets from the Target-Store are randomly selected (by two hardware-based secure pseudo-random number generators) as the candidate replacement objects. By comparing the number of valid tags in the set pointed to by the RPTR of each target, we choose the target with a higher count of valid tags for replacement. The algorithmic process is illustrated in \autoref{a2}.

\begin{algorithm}[!t]
\footnotesize
\caption{Load-balancing replacement algorithm}\label{a2}
\LinesNumbered
\KwIn{{\tt Rand$_0$}, {\tt Rand$_1$}, {\tt BTB\_State}, \textit{Targets}}
\KwOut{{\tt Final\_Target}}

\If{{\tt BTB\_State} $==$ {Miss}}{

{\tt Target$_0$} $\gets$ \textit{Targets}$\text{[}${\tt Rand$_0$}$\text{]}$;

{\tt Target$_1$} $\gets$ \textit{Targets}$\text{[}${\tt Rand$_1$}$\text{]}$;


{\tt M$_0$ $\gets$} Number of valid tags in the set pointed to by the RPTR of {\tt Target$_0$};

{\tt M$_1$ $\gets$} Number of valid tags in the set pointed to by the RPTR of {\tt Target$_1$};

\uIf{{\tt M$_0$} $>=$ {\tt M$_1$}}{
    {\tt Final\_Target} $\gets$ {\tt Target$_0$};
}\Else{
    {\tt Final\_Target} $\gets$ {\tt Target$_1$};
}}

\end{algorithm} 

With the aforementioned mechanism, the occurrence of DE is contingent upon the case that both selected sets are devoid of invalid tags. The proposed load-balancing replacement algorithm significantly promotes a balanced distribution of invalid tags across sets, effectively reducing the probability of DE. The corresponding analysis and discussion will be detailed in \autoref{lab:s}.

\subsection{Discussion of CIBPU}

\mypara{Key Management in CIBPU}
In CIPHT or CIBTB, all the keys required are derived from the Address thread ID (TID) and PC through a hardware encryption function, making them imperceptible to any software, including the operating system. Since the conflict information in CIPHT or CIBTB cannot be perceptible to attackers, all the mapping-related information remains undisclosed. As a result, CIPHT or CIBTB does not require key updates. We suggest employing a hardware-based implementation of a physically unclonable function (PUF))~\cite{zhang2019physical} for key generation. Additionally, for generating all random numbers in CIPHT or CIBTB, we propose using a hardware-based secure random number generator~\cite{cox2011intel}.

\mypara{Parameter selection in CIBPU} The parameter selection principle in CIBPU is to minimize hardware overhead while ensuring security during the system life cycle. 
Among the BPU sizes in currently common CPUs, The number of skews for CIPHT is fixed to three, and CIBTB has two is the most reasonable choice, and the number of additional tags in CIBTB needs to be determined according to the specific size of the BTB. See~\autoref{sec:Security} for detailed security analysis.

\section{Security Analysis}
\label{sec:Security}

In this section, we will perform a detailed analysis of the impact of the two types of attacks mentioned in \autoref{sec:back} on the proposed CIPHT and CIBTB. 
Due to the fact that real-world BPU side/covert-channel attacks typically rely on the attacker's ability to use fine-grained control over the execution of victim processes, such as through the use of Intel SGX's single-stepping framework, and given that the implementation of CIBPU depends on open-source simulators gem5~\cite{gem5simulatorsystem} and the CPU core SonicBOOM~\cite{SonicBOOM3rdGeneration2020}, which do not possess the capability to reproduce such conditions, we prove the security of CIBPU more comprehensively and fundamentally through mathematical modeling rather than attack simulations.
The symbols frequently used in our analysis are provided in \autoref{tab:2}.


\begin{table}[!t]
  \caption{Symbols frequently used in this paper.}
  \centering
    \begin{tabular}{clc}
\midrule[0.8pt]
    \multicolumn{1}{c}{\textbf{Symbol}} & 
    \multicolumn{1}{c}{\textbf{Meaning}} & 
    \multicolumn{1}{c}{\textbf{Default}} \\
\midrule[0.8pt]
    $A_{\text{pht}}$     & \makecell[l]{Number of attacks on PHT} & / \\  
    $A_{\text{btb}}$     & \makecell[l]{Number of attacks on BTB} & / \\ 
    $I_{\text{pht}}$     & \makecell[l]{Number of bits of PHT index} & 13 \\  
    $I_{\text{btb}}$     & \makecell[l]{Number of bits of BTB index} & 12\\  
    $T_{\text{pht}}$     & \makecell[l]{Number of bits of PHT tag} & 12 \\  
    $T_{\text{btb}}$     & \makecell[l]{Number of bits of BTB tag} & 12 \\  
    $N_{\text{btb}}$     & \makecell[l]{Number of bits of BTB target} & 48 \\  
    $W$     & \makecell[l]{Capacity of a bin (Number of BTB ways)} & 8  \\ 
    $N_{\text{bin}}$ & \makecell[l]{Number of bins (Number of BTB sets)} & 4K \\ 
    $N_{\text{ball}}$ & \makecell[l]{Number of balls (Number of BTB targets)}& 4K $\times$ 8 \\  
    $P_N$   & \makecell[l]{Probability that a bin has $N$ balls} & /  \\  
    $P_{n \leq N}$ & \makecell[l]{Probability that a bin has $\leq N$ balls} & / \\  
    $P_{n \geq N}$ & \makecell[l]{Probability that a bin has $\geq N$ balls} & /  \\  
    $P_{X \Rightarrow Y}$ & \makecell[l]{Probability that a bin with \\ $X$  balls transitioning to $Y$ Balls} & /  \\
    $P_{N_{\text{est}}}$ & \makecell[l]{Estimated probability that a bin has $N$ balls } & /  \\
    $P_{N_{\text{obs}}}$ & \makecell[l]{Observed probability that a bin has $N$ balls } & /  \\
\midrule[0.8pt]

    \end{tabular}%
  \label{tab:2}%
\end{table}

\subsection{Analysis on Reuse-based Attacks}

In BPU, this type of attacks require redirecting the victim process's set to a set controlled by the attacker. The attacker can achieve this by controlling multiple sets. On the other hand, the attacker must successfully hit BTB or PHT, which requires the attacker's and victim's tags to be exactly the same with a probability of $\frac{1}{2^{T_{\text{pht}}}}$. We will now calculate the theoretical number of attack iterations required separately for CIPHT and CIBTB.

\mypara{Reuse-based attacks on CIPHT} 
We assume that the attacker has the capability to filter out the mutual influence between their own branches. In the scenario where the attacker does not know the location of the victim's branch, due to the existence of the index encryption mechanism, content encryption mechanism (encrypting the tag and the stored 2-bit state information), and our protection machanism, the average number of attempts for the attacker to launch a successful attack can be calculated 
as
\begin{align}\label{e1}
A_{\text{pht}}=(2^{I_{\text{pht}}}\times 2^{T_{\text{pht}}}\times 2^2)^3~,
\end{align}
where $A_{\text{pht}}$, ${I_{\text{pht}}}$ and $T_{\text{pht}}$ are explained in \autoref{tab:2}. By substituting the default parameters into \autoref{e1}, we can calculate the theoretical estimated number of attacks required by the attacker to be approximately 2.4$\times$10$^{\text{24}}$ (taking more than 10$^{\text{4}}$ years).

\mypara{Reuse-based attacks on CIBTB} 
Similarly, in our BTB structure, due to the existence of the index encryption mechanism and content encryption mechanism (encrypting the tag and target), the average number of attempts for the attacker to launch a successful attack can be calculated as
\begin{align}\label{e2}
A_{\text{btb}}=2^{I_{\text{pht}}}\times 2^{T_{\text{pht}}}\times 2^{N_{\text{btb}}}~.
\end{align}

By substituting the default parameters into \autoref{e2}, we can calculate that the theoretical value of the required number of attack attempts for the attacker is approximately 5$\times$10$^{\text{21}}$ (taking more than 10$^{\text{7}}$ {years}).

\subsection{Analysis on Eviction-based Attacks}
\label{lab:s}

Among the eviction-based attacks launched on the conventional BTB structure, an attack like Jump~\cite{evtyushkin2016jump} can observe the eviction of its own entry in the BTB, and thus obtain the address of the branch executed by the victim. Attackers can extend the attacks against random caches to BTB. These attacks adopt the state-of-the-art techniques for finding eviction sets, such as Prime+Prune+Probe (PPP)~\cite{purnal2021systematic}, Group-Elimination Method (GEM)~\cite{qureshi2019new} and even a more advanced method for finding eviction sets proposed in the future. 


We make the following assumptions for security analysis:

\begin{itemize}
    \item We assume an idealized encryption algorithm that can randomly map a given row to an arbitrary location in memory with equal probability (the set mapping function is completely random and the key is secret). This ensures that the PC is uniformly mapped to Tag-Store in a manner unknown to the attacker, thus not directly causing DE. Furthermore, we assume that the maps of two BTB skews (generated with different keys) are independent.
    \item While the attacker program is looking for the eviction set, no other programs are running. If this is not the case, a third-party program will access the BPU, which will affect the attacker's theoretical time to find the eviction set. In most realistic cases, the attacker can only achieve a fraction of the theoretical number of times needed to find the eviction set, causing the attack to fail.
    \item Due to the great progress of the latest research, the number of accesses required to build an eviction set has been reduced. The most advanced research~\cite{liu2015last,qureshi2019new} shows that at least several hundreds of DEs are needed to build an eviction set. We define two levels of security assurance. The first level is considered insecure if any instance of DE occurs. We take a conservative approach by assuming that attackers can potentially obtain sensitive information with just one DE event. The second level represents the security level adopted by most SBPU schemes~\cite{zhao2022hybp,zhang2022stbpu}, where the discovery of the eviction set for a specific target is considered insecure.

\end{itemize}

To demonstrate how eviction-based attacks can be eliminated, next we estimate the incidence of DE for conventional BTB structures, and explain why a DE is unlikely to occur for CIBTB during the lifecycle of the system.

\mypara{Bins-and-balls model} 
For eviction-based attacks, we  model the operation of attackers using the bins-and-balls problem~\cite{luczak2005power}. Specifically, we represent the attacker's accesses to BTB as the number of balls thrown, and the number of sets in BTB as the number of bins. In the case where the attacker is unaware of the BTB indexing pattern, the act of accessing BTB is analogous to randomly throwing a ball into one of the bins. The scenario where the attacker is mapped to the same set as the victim is modeled as throwing a ball into the same bin. When a ball is thrown into a bin that is already full, it causes one ball to overflow from that bin, which corresponds to an eviction when a branch instruction accesses BTB. We consider the number of Tag-Store capacity of each set as the capacity of the bins, which is equal to $W$ in conventional BTB structure. The total number of entries in the Target-Store of BTB represents the total number of balls that can be thrown. It is important to note that in the conventional BTB structure, the number of tag entries in the Tag-Store is equal to the number of data entries in the Target-Store. However, in CIBTB, the number of tag entries in the Tag-Store is greater than the number of data entries in the Target-Store.

\mypara{Eviction-based attacks on conventional BTB} 
The probability $p$ that an attacker's ball lands in the same bin as the victim's ball is given as
\begin{align}\label{equal1}
p = 2^{-I_{\text{btb}}} = \frac{1}{N_{\text{bin}}}~.
\end{align}
Let $L$ denote the number of attacks (balls thrown). As the balls are continuously thrown ($L$ increases gradually), the probability of having $N$ balls in the bin where the victim's ball resides follows a binomial distribution, which is
\begin{align}\label{equal2}
P_N = \tbinom{L}{N}(1-p)^{L-N}\times p^N~.
\end{align}
Combining \autoref{equal1} and \autoref{equal2}, as $L$ increases, it is observed that $N$ becomes larger while $p$ remains small. Therefore, $P_N$ approximately follows a Poisson distribution, which
\begin{align}\label{equal3}
P_N =\frac{e^{-\lambda}\times \lambda^N}{N!} , \quad (\lambda = \frac{L}{N_{\text{bin}}})~.
\end{align}
When the expected value of encountering a DE is 0.5, which means that the average number of a successful eviction by the attacker in accessing BTB is approximately half of the number of attempts required for guaranteeing a successful eviction, $L$ satisfies
\begin{align}\label{equal4}
N_{\text{bin}}\times P_{(W+1)} = N_{\text{bin}}\times \frac{e^{-\frac{L}{N_{\text{bin}}}}\times (\frac{L}{N_{\text{bin}}})^{W+1}}{(W+1)!}=0.5~.
\end{align}

Under the baseline parameters, we can calculate that the estimated theoretical result is approximately $L^1_{\text{est}} \approx 7690$, which represents the number of theoretical attack attempts required to break the first level of security. Using the bins-and-balls model, we randomly throw balls into bins and observe the number of balls thrown when the bin contains $(W+1)$ balls for the first time. We set 10 trillion balls in our experiment, and the average result was approximately $L^1_{\text{obs}} \approx 7730$.

For the second level of security, attackers can use the state-of-the-art algorithm GEM to construct every eviction set. The evaluation of security is conducted using GEM because, without partitioning and randomization structures, the efficiency of bottom-up strategies like PPP would be compromised. The original configuration of GEM~\cite{qureshi2019new} is employed, where the attacker sets the group size as $G=W+1$. The number of BTB accesses to construct the eviction set using GEM is approximately $L^2_{\text{est}} = 2.3 \times W \times L^1_{\text{est}} \approx 1.4 \times 10^5$  (taking about 0.14 ms).

In summary, to protect the conventional BTB structure from eviction attacks, it would be necessary to re-randomize the indexing scheme within a time frame corresponding to the attacker's access count at the magnitude of $10^6$ (taking about 1 ms).

\mypara{Eviction-based attacks on CIBTB} 
Similar to MIRAGE~\cite{saileshwar2021mirage}, the number of balls in a bin can both increase and decrease. Each bin follows a birth-death chain~\cite{lilja2005measuring}, which is a type of Markov chain where the state variable (number of balls in the bin) increases or decreases by one due to birth or death events (insertion or deletion), respectively. We can leverage the classical result of the birth-death chain, which states that in a steady state, the probability of each state converges to a stable value, and the net transition rate between any two states is zero. By equating the probability of a bucket with $N$ balls transitioning to $N+1$ balls and vice versa, we have
\begin{align}\label{equal5}
P_{N\Rightarrow N+1} = P_{N+1\Rightarrow N}~.
\end{align}
We note that a bin with $N$ balls transitioning to $N+1$ balls on a ball insertion if any of the following conditions is satisfied:
\begin{itemize}
\item[1.] The selected bins from both \textit{Skew}-0 and \textit{Skew}-1 have $N$ balls.
\item[2.] The selected bin from \textit{Skew}-0 has $N$ balls and that from \textit{Skew}-1 has more than $N$ balls.
\item[3.] The selected bin from \textit{Skew}-1 has $N$ balls and that from \textit{Skew}-0 has more than $N$ balls.
\end{itemize}
Therefore, $P_{N\Rightarrow N+1}$ satisfies
\begin{align}\label{equal6}
P_{N\Rightarrow N+1} =  P_N^2 + 2\times P_NP_{n\geq N+1}~.
\end{align}
A bin with $N+1$ balls transitions to $N$ balls on a ball insertion if any of the following conditions is satisfied:
\begin{itemize}
\item[1.] The bins pointed to by both \textit{Ball}-0 and \textit{Ball}-1 have $N+1$ balls.
\item[2.] The bin pointed to by \textit{Ball}-1 has $N+1$ balls and by \textit{Ball}-1 has less than $N+1$ balls.
\item[3.] The bin pointed to by \textit{Ball}-2 has $N+1$ balls and by \textit{Ball}-0 has less than $N+1$ balls.
\end{itemize}
Therefore, $P_{N+1\Rightarrow N}$ satisfies
\begin{align}\label{equal7}
    \begin{split}
         P_{N+1\Rightarrow N} = & \left[\frac{(N+1)N_{\text{ball}}}{N_{\text{bin}}}P_{N+1}\right]^2 \\
         +&2\times \frac{(N+1)·N_{\text{ball}}}{N_{\text{bin}}}P_{N+1}\times 
        \sum_{i=0}^N\frac{i\times N_{\text{ball}}}{N_{\text{bin}}}P_i~.
    \end{split}
\end{align}
Combining \autoref{equal5},~\autoref{equal6} and~\autoref{equal7}, and substituting the default values from \autoref{tab:2}, we have
\begin{align}\label{equal8}
    \begin{split}
        \frac{(N+1)^2}{64}P_{N+1}^2 
        & +\left(\frac{N+1}{4}\sum_{i=0}^N\frac{i}{8}P_i\right)P_{N+1}\\
        & -\left[P_N^2+2P_N\left(1-\sum_{i=0}^{N}P_i\right)\right]= 0~.
    \end{split}
\end{align}
As $n$ grows, $P_N\rightarrow 0$, $\sum_{i=0}^NP_i\rightarrow 1$ and $1- \sum_{i=0}^NP_i \ll P_N$. The approximate value of $P_{N+1}$ is
\begin{align}\label{equal9}
P_{N+1} = \frac{P_N^2}{\frac{N+1}{4}\sum_{i=0}^N\frac{i}{8}P_i}~.
\end{align}

\begin{figure}[t]
    \centering
    \includegraphics[width=0.8\linewidth]{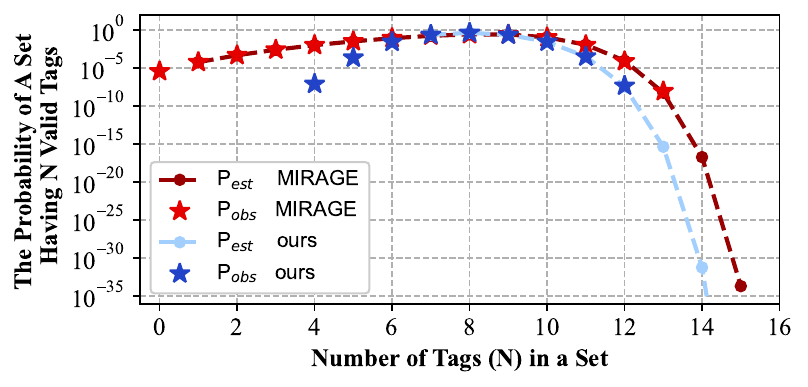}
    \vspace{-1.em}
    \caption{Probability of a set having $N$ valid tags - Estimated analytically ($P_{\text{est}}$) and Observed ($P_{\text{obs}}$).} 
    \label{figs1}
    \vspace{-1em}
\end{figure}

By conducting a ball-dropping experiment with 10 trillion balls, we obtain the probability of a bin with $N \in [4, 12]$ balls, denoted as $P_{N_{\text{obs}}}$ (when $N < 4$ or $N > 12$, we do not observe any bins with those numbers of balls loaded). 
Substituting the ${P_{4_{\text{obs}}}}$ and ${P_{5_{\text{obs}}}}$ value into \autoref{equal9}, we recursively calculate $P_{\text{est}}$ for $N \in[6, 14]$. \autoref{figs1} shows the empirical observations ($P_{\text{obs}}$) and the analytical estimation ($P_{\text{est}}$) probabilities for a bin with $N$ balls. It can be observed that our experimental values align perfectly with the curve of the theoretical formula.

\autoref{figs1} shows that, similar to MIRAGE~\cite{saileshwar2021mirage}, the probability of a set having $N$ tags decreases double-exponentially beyond 8 tags per set. The theoretical and experimental results from the MIRAGE scheme are also presented in the figure. We compared the results of both schemes. For $N = 12/13/14$, the probability reaches $10^{-5}/10^{-9}/10^{-17}$ in MIRAGE and reaches $10^{-8}/10^{-15}/10^{-31}$ in CIBTB.

\begin{figure}[!t]
    \centering
    \includegraphics[width=0.8\linewidth]{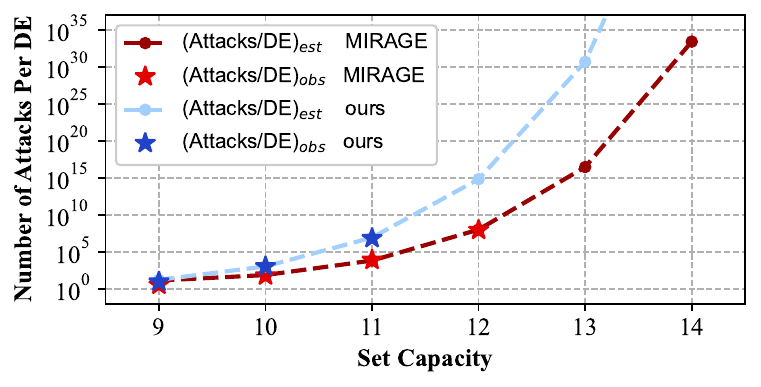}
    \vspace{-1em}
    \caption{Frequency of DE, as set-capacity
varies – both analytically estimated (Attacks/DE)$_{\text{est}}$ and empirically observed (Attacks/DE)$_{\text{obs}}$ results are shown.} 
    \label{figs2}
    \vspace{0em}
\end{figure}

Using these probabilities, we can estimate the frequency of DE. For a bin of capacity $W$, the DE-probability is the probability that a bin with $W$ balls gets to $W+1$ balls. By setting $N = W$ in \autoref{equal9}, we get the DE-probability as
\begin{align}\label{equal10}
P_{\text{DE}} = \frac{4\times P_{W+1}^2}{(W+2)\sum_{i=0}^{W+1}\frac{i}{8}P_i}~.
\end{align}
\autoref{figs2} shows the estimated values (denoted as (Attacks/DE)$_{\text{est}}$) of the average number of attacks required for a single DE event calculated using \autoref{equal10}, as well as the observed values (denoted as (Attacks/DE)$_{\text{obs}}$) obtained from the simulation with 10 trillion balls. It can be observed that our experimental values align perfectly with the theoretical curve. For $W = 13$, a DE occurs every $5\times 10^{31}$ attacks with CIBTB while an DE occurs every $7\times 10^{16}$ attacks with MIRAGE. Therefore, our approach can guarantee system security throughout its lifecycle with a smaller $W$ compared to MIRAGE, which significantly reduces hardware overhead. In conclusion, CIBTB ensures first-level security by adding an additional 5 tags to each set on top of the 8-way BTB structure.

\subsection{Security Summary}

According to the above analysis, the number of attack attempts required by an attacker to launch different types of attacks against CIBPU is summarized in \autoref{tab3}. Based on the analytical results, one can find the proposed scheme ensures strong security throughout the system's lifecycle. 
This is precisely why CIBPU doesn't require key re-randomization.
On the other hand, since DE doesn't occur throughout the system's lifecycle, it doesn't result in effective information leakage. This, in turn, prevents any possibility of successfully reversing the encryption method of CIBPU.



\begin{table}[!t]
\caption{Number of attack attempts and time required for different types of attacks.\label{tab3}}
\centering

\begin{tabular}{lcc}
\midrule[0.8pt]
        \multicolumn{1}{c}{\textbf{Attack Type}} & 
        \multicolumn{1}{c}{\textbf{PHT}} & 
        \multicolumn{1}{c}{\textbf{BTB}}\\
\midrule[0.8pt]
        Reuse-based Attack & \makecell[c]{$2.4\times10^{24}$ \\ (\textgreater $10^{4}$ years)} & \makecell[c]{$5\times10^{21}$ \\ (\textgreater $10^{7}$ years)} \\
\midrule
        Eviction-based Attack & / & \makecell[c]{$5\times10^{31}$ \\ (\textgreater $10^{14}$ years)} \\
\midrule[0.8pt]
\end{tabular}
\vspace{-5pt}

\end{table}

\section{Evaluation}\label{sec:Evaluation}
\subsection{Methodology}\label{sec:A}

We utilize the cycle-accurate microarchitecture simulator gem5 with the DerivO3CPU model configuration to implement the proposed scheme and the state-of-the-art schemes for comparison. This configuration represents an out-of-order execution processor setup that closely resembles the architecture of contemporary commercial processors. Furthermore, it supports both single-threaded execution and SMT mechanism, thereby facilitating our performance evaluation in both scenarios.

On the other hand, we also implement CIBPU in the RISC-V open-source processor, SonicBOOM. The complete processor is burned upon the KC-705~\cite{fpga} FPGA evaluation board. By building such an SoC verification platform, we evaluate the performance of the processor under the CIBPU scheme, including peripherals such as serial ports, Ethernet ports, and an SD card controller. A Linux operating system is run based on the RISC-V architecture in this platform. The SD card contains the RISC-V Open Source Supervisor Binary Interface (OpenSBI), U-Boot, Linux kernel, and Debian root FS. The configuration parameters of gem5 and SonicBOOM are shown in \autoref{tab:para}. 

\begin{scriptsize}
\begin{table}[!t]
\caption{gem5 and SonicBOOM parameters.\label{tab:para}}
\centering
\begin{tabular}{lll}
\midrule[0.8pt]
        \multicolumn{1}{l}{\textbf{Parameter}} & 
        \multicolumn{1}{c}{\textbf{gem5}} & 
        \multicolumn{1}{c}{\textbf{SonicBOOM}}\\
\midrule[0.8pt]
        ISA & \makecell[l]{Single thread: RISC-V \\ SMT: x86 SMT-2}& \makecell[l]{Single thread: RISC-V \\ SMT: Not support }\\
\midrule
        Frequency & 5GHz & FPGA@50Mhz \\
        CPU type & \makecell[l]{DerivO3CPU \\ 8-decode 8-issue}  & \makecell[l]{Medium-Out-of-Order \\ 2-decode 4-issue}\\
\midrule
        BTB & 8K, 8-way & 8K, 8-way\\
        PHT & TAGE\_SC\_L: 4K & TAGE: 4K\\
        L1 ICache & 32KB, 8-way, 64B line & 16KB, 4-way, 64B line\\
        L1 DCache & 32KB, 8-way, 64B line & 16KB, 4-way, 64B line \\
        L2 Cache  & 1MB, 16-way, 64B line & 512KB, 16-way, 64B line \\
        L3 Cache  & 4MB, 32-way, 64B line & Not support \\
\midrule[0.8pt]
\end{tabular}
\vspace{-5pt}
\end{table}
\end{scriptsize}

\subsection{Performance Evaluation with gem5}
CIBPU is compared with the state-of-the-art schemes (listed in \autoref{tab:1}) in gem5 by executing SPEC2017 benchmarks. Specifically, the schemes used for comparison include:

\begin{figure}[!t]
    \centering
    \includegraphics[width=0.8\linewidth]{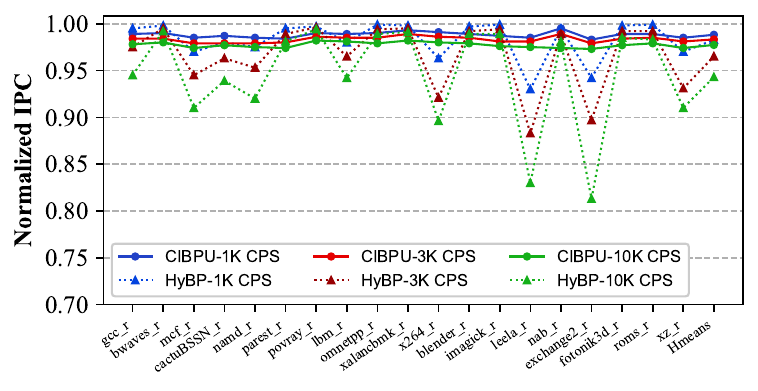}
    \vspace{-1em}
    \caption{Normalized IPC of HyBP and ours on a single-threaded core by different frequency of context switch. The data is normalized to the data of baseline of 1K CPS.} 
    \label{fig:eva1}
    \vspace{-15pt}
\end{figure}

\begin{itemize}
    \item \textbf{Baseline:} The original DerivO3CPU using TAGE-SC-L PHT and basic BTB without any defense mechanism. 
    \item \textbf{Flush:} Flushing the predictor completely upon context switches or privilege changes.
    \item \textbf{Partition:} Partitioning PHT and BTB for each thread and partitioning each table among thread/privilege.
    \item \textbf{Replication:} Replicating PHT and BTB for each thread and partitioning each table among thread/privilege.
    \item \textbf{HyBP:} The original DerivO3CPU using TAGE-SC-L PHT and BTB with encryption-based protection mechanism. The index and content keys of BTB and PHT are changed during context switches.
    \item \textbf{STBPU:} The original DerivO3CPU using TAGE-SC-L PHT and BTB with encryption-based protection mechanism. The index and content keys of BTB and PHT are changed when the number of detected malicious BPU events reaches a threshold.
\end{itemize}

We simulate each scheme across 19 SPEC2017 workloads and calculate the Harmonic means (Hmeans)~\cite{michaud2012demystifying} of instructions per cycle (IPC) since each workload is equally valued. All simulations are performed by executing 110 million instructions after a warm-up of 10 million instructions.

\mypara{Performance evaluation on single-threaded core} 
We observe that certain applications in SPEC2017 are highly sensitive to the historical information stored in the branch predictor, while the branch history preservation capability of most comparative schemes is highly influenced by the frequency of context switches. In a Linux operating system, thousands to tens of thousands of context switches are typically occurring within a 1-second time frame. We evaluate the performance of different schemes at a CPU with a main frequency of 5 GHz while varying the context switch frequencies. 
\autoref{fig:eva1} illustrates the performance impact on a single application under different context switch frequencies for two representative protection schemes: HyBP and our proposed scheme. The unit of the number of context switches per second is “CPS” (context switches per second). It can be observed that under very frequent context switches, certain programs in the HyBP scheme may experience a performance decrease of nearly 19\%. Under different context switching frequencies, our proposed scheme exhibits an average performance degradation ranging from 1.12\% to 2.20\%, while HyBP experiences an average performance loss ranging from 1.73\% to 5.69\%.

\begin{figure}[t]
    \centering
    \includegraphics[width=0.8\linewidth]{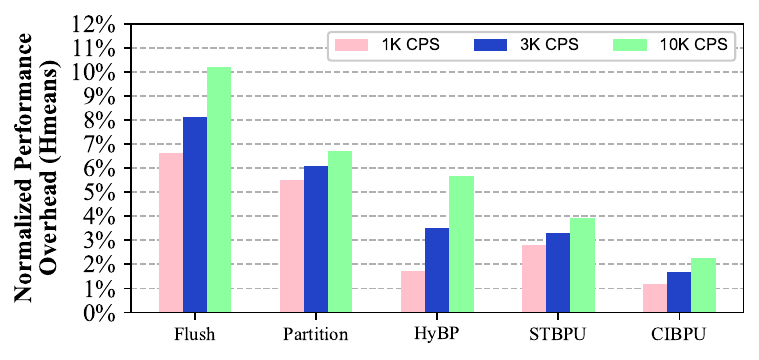}
    \vspace{-1em}
    \caption{Performance degradation of different schemes for a single-threaded core under different frequency of context switch. The data is normalized to the data of baseline of 1K CPS.} 
    \label{fig:eva2}
    \vspace{-15pt}
\end{figure}

Next, we compare the CIBPU scheme with other schemes. \autoref{fig:eva2} illustrates the average performance degradation of different schemes at varying CPS. It is worth noting that the Replication scheme does not affect system performance under single-threaded operation. Therefore, we do not test this scheme. We observe that our CIBPU scheme consistently incurs a performance loss below 2.20\%, which is lower than other protection schemes. In conclusion, CIBPU exhibits performance advantages in single-threaded mode, and its performance loss is not sensitive to the context switch frequency.

\begin{figure}[t]
    \centering
    \includegraphics[width=0.8\linewidth]{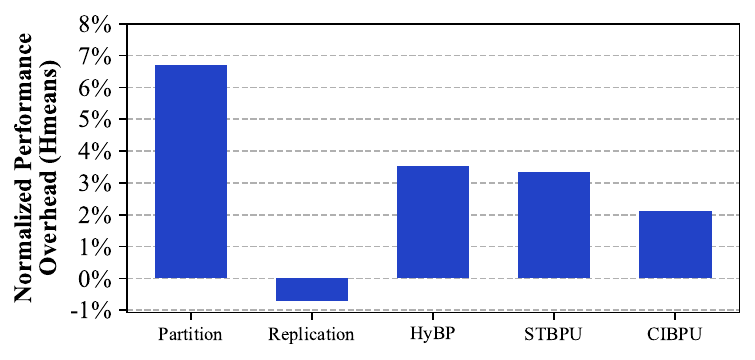}
    \vspace{-.5em}
    \caption{Performance degradation shown as Hmeans of different schemes for an SMT-2 core under different frequency of context switch. The data is normalized to the baseline.} 
    \label{fig:eva3}
    \vspace{-1.em}
\end{figure}

\mypara{Performance evaluation on SMT core} 
The same SBPU schemes are tested except for the Flush scheme, because in the SMT operation mode, the Flush scheme no longer provides any protection capability. In order to focus on testing the performance degradation of different solutions in SMT mode, CPS is fixed to 3000, we employ the standard approach proposed in~\cite{cazorla2004dynamically}, which involves paired workloads and executing them in SMT-2 mode. \autoref{fig:eva3} compares the average performance loss of these paired workloads under different protection mechanisms in SMT mode (evaluated using IPC). The operation in SMT mode is more complex compared to single-threaded mode, as multiple concurrent processes share the BPU resources, which can lead to branch conflicts and consequent performance degradation (notably, this is not malicious behavior). Various protection mechanisms may improve or exacerbate these conflict situations. To ensure fairness, we measure the sum of IPC for all threads and then calculate the Hmeans of this metric for different workloads, normalizing it to the baseline.

Note that the Replication scheme even increases system performance in SMT mode, as it effectively isolates the conflicts in BPU resources between two concurrently running processes. In addition, compared to the other schemes, CIBPU incurs the lowest performance overhead, with an average performance loss of 2.10\%.

\subsection{Hardware Storage Cost Estimation}\label{H}

The main hardware storage overhead of CIBPU contains two parts, i.e., additional PHT entries and additional BTB tag entries along with the extra storage space required in the BTB for storing FPTR and RPTR. Based on the number of entries in the BPU components of the commercial high-performance processor architecture ZEN4~\cite{10067540}, the additional hardware storage overhead introduced by CIBPU is approximately 59.3KB. The additional hardware storage overhead of HyBP consists of two main parts. The first part is the duplication of basic entries for PHT and BTB, and the second part is the storage of keys for different threads and privilege levels. In the case of SMT-2, the total overhead is approximately 39.6KB, while in the case of SMT-8, the total overhead is approximately 197.5KB. The Replication mechanism replicates the BPU entries for each combination of thread-level and privilege-level, resulting in storage overheads of 215.3KB and 1.04MB for processor architectures with SMT-2 and SMT-8, respectively.

\subsection{Analysis on the Comparison of CIBPU with other SBPUs}
Overall, we test the performance overhead of different protection schemes in both single-threaded and SMT modes using the gem5 simulator. \autoref{tab:1} shows the comparison of CIBPU with five state-of-the-art SBPU schemes.


The Flush scheme requires no additional hardware storage cost but involves refreshing BPU information during context or privilege switches. However, it cannot defend against attacks between processes in SMT mode and has significant performance overhead in single-threaded mode, which increases with frequent context switches. 
The Partition scheme cannot defence attacks in single-threaded mode unless each thread is individually partitioned, which is impractical. The presence of partitions reduces the available BPU resources for each thread, leading to relatively higher performance losses in our performance tests. 
In Replication scheme, the presence of BPU isolation between the two concurrently running threads eliminates BPU conflicts, and therefore, in certain BPU-sensitive workloads, it can even improve performance. However, the substantial additional storage overhead makes this scheme almost impractical to implement.

HyBP has made important explorations in balancing security, performance loss, and hardware storage overhead. However, similar to the Flush scheme, the BPU information in HyBP becomes invalid after a context switch since the keys are switched during the context switch. As a result, the performance loss of HyBP increases with more frequent context switches. Additionally, to hide the latency of generating new keys during frequent key switching, HyBP requires pre-computation of keys and additional storage space to store keys for each thread. 
STBPU focuses on malicious events in BPU and automatically remaps the indices and encrypts the contents when the number of BPU evictions and BPU misses reaches a set threshold. The feasibility of STBPU relies on a high level of software-hardware integration. This solution does not offer performance advantages compared to HyBP under low context switch rates.

Our proposed CIBPU achieves a good balance among performance, hardware storage overhead, and security. CIBPU reconstructs the internal structure of BPU to ensure security without the need for key re-randomization. It guarantees security in both single-thread and SMT modes, with the lowest performance overhead among the aforementioned solutions. We provide BTB with additional invalid tags, store extra content for cross-referencing between Tag-Store and Target-Store, and increase the extra content storage in PHT. This requires a limited additional storage overhead, and this overhead does not increase with an increase in the number of concurrent threads. 

\subsection{Validation with RISC-V RTL}

\begin{figure}[!t]
    \centering
    \includegraphics[width=0.8\linewidth]{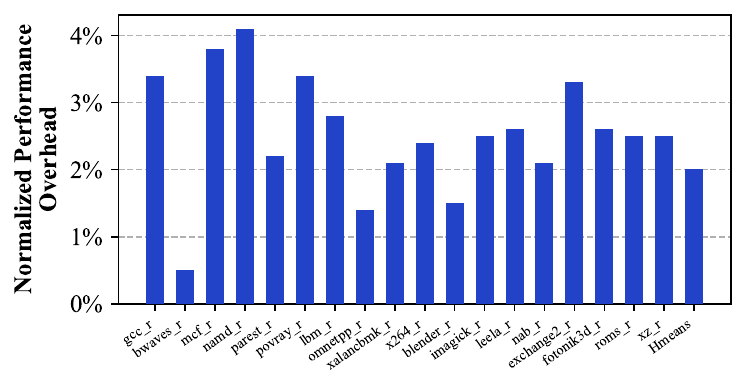}
    \vspace{-1em}
    \caption{Performance degradation of SonicBOOM with CIBPU protection scheme (normalized to the baseline).} 
    \label{fig:fpga}
    \vspace{-0.em}
\end{figure}

\mypara{Performance evaluation based on hardware prototype}
We further implement CIBPU in the open-source RISC-V processor, SonicBOOM, and build peripheral circuits for evaluation in the KC-705 FPGA development board. We run the Linux operating system on the FPGA and execute the SPEC2017 workloads for performance evaluation. Although SonicBOOM does not support SMT mode, the performance evaluation in single-core mode still provides meaningful reference. Specifically, ~\autoref{fig:fpga} illustrates the performance degradation under the configuration specified in ~\autoref{tab:para}. It can be observed that the average performance degradation ($\approx$2.01\%) closely aligns with the results obtained from the gem5 simulator ($\approx$1.12\%--2.10\%).

\begin{table}[!t]
\caption{CIBPU resources utilization.\label{tab:fpga}}
\centering
\begin{tabular}{ccccc}
\midrule[0.8pt]
       & LUTs & FFs & DSPs & BRAMs \\
\midrule[0.8pt]
        Baseline & 169692 & 94784 & 36 & 187 \\
\midrule
        CIBPU & \makecell{176327\\(+3.91\%)} & \makecell{99241\\(+4.70\%)} & 36 & \makecell{198\\(+5.88\%)} \\
\midrule[0.8pt]
\end{tabular}
\vspace{-15pt}
\end{table}

\mypara{FPGA resources utilization}  
We synthesize the actual circuits of the SonicBOOM processor with and without the CIBPU protection mechanism (the latter is used as the baseline) using the Vivado 2021.2 tool. The resource usage of both designs can be obtained from the synthesizing report. As shown in ~\autoref{tab:fpga}, compared to the baseline, CIBPU causes less than 6\% of each type of resource utilization. Its main overhead comes from the construction of the key generator and the additional storage required for the mechanisms. The former is primarily composed of LUTs (Look-Up-Tables) and FFs (Flip-Flops), while the latter is mainly composed of BRAMs (Block RAMs) and FFs (Flip-Flops).

\section{Related Work}

To defend against BPU covert-channel attacks and Spectre attacks using BPU, countermeasures that have been proposed fall into the following two categories:

\subsection{Defenses Based on Controlling Program Execution Flow}

\mypara{Preventing information leakage from cache} 
There have been plenty of countermeasures~\cite{yan2018invisispec, weisse2019nda, li2019conditional, khasawneh2019safespec, saileshwar2019cleanupspec, yu2019speculative} preventing speculative execution from generating visible micro-architectural state. InvisiSpec~\cite{yan2018invisispec} proposes to load an unsafe speculative execution into a speculative buffer instead of loading it into cache. A latest research shows that the invisible-micro-architectural-based defenses have been recently breached by the speculative interference attack~\cite{behnia2021speculative}, which exploits micro-architectural contention on hardware components such as MSHR to leak key information. On the other hand, unXpec~\cite{li2022unxpec} exploited secret-dependent timing differences to break undo-based safe speculation scheme. Another perspective is that covert channels can be eliminated by making the execution time constant~\cite{gruss2016flush+} or by interfering with the access time to interfere with the attacker's measurements~\cite{trilla2018cache}. However, such methods are only applicable to solve the leakage problem of cache covert channel but cannot eliminate the covert channel of the BPU itself.

\mypara{Security checks on software} 
Spectector~\cite{guarnieri2020spectector} and SpecTaint~\cite{qi2021spectaint} statically and dynamically analyze source code and identified vulnerabilities, respectively, and insert the {\it i.fence} instruction in the code fragments identified as insecure to prevent speculative execution. This type of methods relies on recognition accuracy and does not guarantee adequate security. Moreover, like the first type of scheme, it does not have the ability to perceive the BPU covert channel.

\subsection{Defenses Based on Security Enhancement on BPU}

\mypara{SBPU based on physical isolation}
BRB~\cite{vougioukas2019brb} provides a private entire history of the directional predictor for each process which is stored and reloaded when the context is switched. 
Sensitive applications in SGX are advised to provide their own private branch prediction tables~\cite{evtyushkin2018branchscope}. 
These methods effectively mitigate PHT reuse-based attacks such as BranchScope but bring unacceptable hardware overhead due to the need of replicating additional table entries. 
Flushing the entire branch predictor when switching context is an expensive operation~\cite{evtyushkin2016understanding} that introduces significant context switching overhead. 
Moreover, it is not valid for SMT processors, because there are no flush events triggered by context switches between simultaneous threads.

\mypara{SBPU based on encryption}
BSUP~\cite{lee2020securing} first proposes to encrypt the index and content of BPU, but this solution does not work for SMT processors. Noisy-Xor-BP~\cite{zhao2021lightweight} and HyBP~\cite{zhao2022hybp} encrypt BPU index and content by using a key generated from a thread-private random number to achieve logical isolation between threads or privilege levels. These approaches re-generate random key upon context and mode switches. STBPU~\cite{zhang2022stbpu} also uses the index and content encryption mechanism, while monitoring the number of specific BPU events to control the frequency of re-randomization to a reasonable value. In industry, Samsung Exynos~\cite{grayson2020evolution} has implemented content encryption in BTB and RSB with a simple substitution cipher, but it only defenses certain Spectre variants (e.g., Spectre-BTB and Spectre-RSB).

\section{Conclusion}

This paper proposed CIBPU, which provides an effective solution for enhancing the security of BPU. By employing encryption mechanisms and implementing protective designs for both PHT and BTB, CIBPU successfully mitigates information reuse attacks and ensures the integrity of PHT content. The encryption of BTB and the introduction of invalid tags further strengthen the protection against attacks on BTB. The load-balancing index and replacement algorithms, combined with providing more invalid tags, effectively neutralize attackers' ability to exploit branch conflicts for malicious purposes. More importantly, it solves the drawbacks caused by the BPU encryption mechanism requiring periodic key switching. Experimental results demonstrate that CIBPU achieves its security objectives without significantly degrading system performance. This research makes significant contributions to system security and provides a novel approach for future investigations in related fields.

\bibliographystyle{IEEEtran}
\bibliography{main}






\vfill

\end{document}